\documentclass[12pt]{article}

\baselineskip=\normalbaselineskip

\usepackage{mathrsfs,amsbsy,amssymb,latexsym,amsfonts,amsmath,amsthm}
\usepackage[nosort]{cite}

\usepackage{bm}
\usepackage{graphicx} 
\allowdisplaybreaks[1]

\makeatletter
\catcode`\@=11
\@addtoreset{equation}{section}

\def\@seccntformat#1{\csname the#1\endcsname.~~}
\makeatother

\begin{document}
\begin{titlepage}
  \renewcommand{\thefootnote}{\fnsymbol{footnote}}
  \vspace*{1.0cm}

\begin{center}
  {\Large \bf
    Comment on the subtlety of defining real-time path integral
    in lattice gauge theories
  }
  \vspace{1.0cm}

  \centerline{
    {Nobuyuki Matsumoto}
    \footnote{E-mail address: 
      nobuyuki.matsumoto@riken.jp}
  }

  \vskip 0.8cm
  {\it RIKEN/BNL Research center, Brookhaven National Laboratory,
    Upton, NY 11973, USA}
  \vskip 1.2cm

  \end{center}

  \begin{abstract}
    Recently, 
    Hoshina, Fujii, and Kikukawa
    pointed out that
    the naive lattice gauge theory action
    in Minkowski signature does not result in
    a unitary theory in the continuum limit,
    and Kanwar and Wagman
    proposed alternative lattice actions
    to the Wilson action without divergences.
    We here show that the subtlety can be understood from
    the asymptotic expansion of the modified Bessel function,
    which has been discussed for path integral of
    compact variables in nonrelativistic quantum mechanics.
    The essential ingredient for defining the appropriate continuum theory 
    is the $i\varepsilon$ prescription,
    and with the proper implementation of the $i\varepsilon$
    we show that the Wilson action can be used
    for the real-time path integrals.
    It is here important that the $i\varepsilon$ should be implemented
    for both timelike and spacelike plaquettes.
    We also argue the reason why
    the $i\varepsilon$ becomes required for the Wilson action
    from the Hamiltonian formalism.
    The $i\varepsilon$ is needed to
    manifestly suppress the contributions from singular paths,
    for which the Wilson action can give different values
    from those of the actual continuum action.
  \end{abstract}
\end{titlepage}

\pagestyle{empty}
\pagestyle{plain}

\setcounter{footnote}{0}

\section{Introduction}
\label{sec:intro}

Real-time path integral \cite{Feynman:1948ur}
has recently been revisited
both analytically
\cite{Witten:2010cx,Tanizaki:2014xba,Turok:2013dfa}
and numerically
\cite{Berges:2006xc, Berges:2007nr, Alexandru:2016gsd,
  Alexandru:2017lqr, Mou:2019tck, Takeda:2021mnc,
  Fujisawa:2021hxh, Asante:2021phx}
for the interest of real-time dynamics in quantum theories.
Especially in the numerical side,
many developments have been made to tame
the infamous sign problem
(e.g., complex Langevin
\cite{Parisi:1983cs,Klauder:1983sp,
  Berges:2006xc,Berges:2007nr,
  Aarts:2009dg,
  Nishimura:2015pba},
contour deformation techniques including Lefschetz thimble methods
\cite{Witten:2010cx, Cristoforetti:2012su, Cristoforetti:2013wha,
  Fujii:2013sra, Tanizaki:2014xba,
  Alexandru:2015sua, Alexandru:2016gsd,
  Fukuma:2017fjq,
  Alexandru:2017oyw, Alexandru:2017lqr,
  Mou:2019tck, Fukuma:2019wbv, Fukuma:2019uot,
  Fukuma:2020fez, Fukuma:2021aoo, Mori:2017pne, Mori:2017nwj, Alexandru:2018fqp,
  Kanwar:2021tkd, Fujisawa:2021hxh},
and tensor renormalization group
\cite{Niggemann:1997cq, Verstraete:2004cf, Levin:2006jai,
  Xie:2009zzd, Gu:2010yh, Xie:2012zzz, Adachi:2019paf, Kadoh:2019kqk, Fukuma:2021cni,
  Takeda:2021mnc, Hirasawa:2021qvh}),
which can enables us to investigate real-time quantum systems
via numerical calculation.
It is thus becoming not only of theoretical interest
but also of practical importance
to establish an appropriate way to calculate
the real-time path integrals.
Recently, Hoshina, Fujii, and Kikukawa \cite{Hoshina:2020gdy}
pointed out that
the naive lattice gauge theory action
in Minkowski signature does not result in
a unitary theory in the continuum limit,
and Kanwar and Wagman \cite{Kanwar:2021tkd}
proposed alternative lattice actions
to the Wilson action removing divergences
to give a well-defined continuum limit.%
\footnote{
  See also \cite{Fatollahi:2016cxk}
  for a discussion on unitarity of
  the time evolution operator and
  the role of imaginary time
  in theories with compact variable.
}
In this paper,
we point out that the subtlety can be understood from
the asymptotic expansion of the modified Bessel function,
which has been discussed in nonrelativistic quantum mechanics
of compact variables \cite{Inomata1979, Junker1987}.
To get rid of the
unwanted
part of the asymptotic expansion,
we need to incorporate the $i\varepsilon$ prescription,
i.e., an infinitesimal Wick rotation \cite{Wick:1954eu}.
The first point of this paper is that
we can use the Wilson action for the numerical studies,
but with the $i\varepsilon$ implemented.
It is also possible to expand the Boltzmann weight
with the characters to define the real-time action,
in which case we express the characters
not with the modified Bessel functions themselves
but with its asymptotic expansion in $\varepsilon \rightarrow +0$.
In the latter case, the $i \varepsilon$ is already built in the action,
and thus, safely setting $\varepsilon=0$, the theory
has an appropriate continuum limit.

In the above two actions,
the key ingredient is the $i\varepsilon$,
which we know is essential in the continuum theory
to obtain the causal structure of the Green functions. 
However, it may seem uncertain why such $i\varepsilon$ is required
without knowing the actual continuum quantum theory.
As the second point of this paper,
starting from the Hamiltonian formalism,
we argue the reason why
the $i\varepsilon$ becomes required for the Wilson action.
Note that, the Wilson action is only guaranteed
to reproduce the action values of the continuum action
for smooth field configurations.
Here, the $i\varepsilon$ is needed to
manifestly suppress the contributions from these singular paths
in the path integral.%
\footnote{
  The author is sincerely grateful to Yoshio Kikukawa and the referee of
  Progress of Theoretical and Experimental Physics
  for pointing out the misstatements
  in the first version of the manuscript
  that was lead by not recognizing
  the well-defined distributional meaning of the Feynman kernel.
  Major part of section~\ref{sec:explanation}
  is revised accordingly from the first version.
}

As an illustrative example,
we begin with a simple
one-dimensional quantum mechanical system in a periodic box \cite{Kanwar:2021tkd}.
We define the lattice action by discretizing time direction 
resulting in a $U(1)$ theory,
and review the subtlety in defining continuum limit of the real-time path integral
for this model
\cite{Inomata1979,Junker1987,Kanwar:2021tkd}.
We in particular explain how the correct continuum limit emerges
with the $i \varepsilon$
by analyzing the asymptotic expansion of the modified Bessel function \cite{Inomata1979}.
We then argue the meaning of the $i\varepsilon$
by deriving the path integral from the Hamiltonian formalism.
This model gives the essential structure for the necessity of the $i\varepsilon$.

With the detailed picture in quantum mechanics,
the lattice gauge field theory can be seen in a straightforward manner.
We first describe the subtlety of real-time path integral
in gauge theories \cite{Kanwar:2021tkd} with the modified Bessel function.
The expansion of the Boltzmann weight with characters
shows that we need to incorporate the $i\varepsilon$
both for the timelike and spacelike plaquettes.
Next, we exemplify that the Wilson action can be used with the $i\varepsilon$
by using the two-dimensional $SU(2)$ and $SU(3)$ theories.
Lastly, we
argue the meaning of the $i\varepsilon$ from the Hamiltonian formalism,
in particular considering the $SU(2)$ Wilson theory \cite{Kogut:1974ag}.

The remaining part of this paper is organized as follows.
In section~\ref{sec:qm},
we first review the subtlety of real-time path integral
in the quantum mechanics
on $S^1$.
We then argue the meaning of the $i\varepsilon$
by deriving the path integral
expression
from the Hamiltonian formalism.
In section~\ref{sec:field_theory}, we move to the lattice gauge theory case.
After describing the subtlety of the real-time path integral similarly to section~\ref{sec:qm},
we demonstrate that the Wilson action can be used with the $i\varepsilon$.
Lastly, we clarify the meaning of the $i\varepsilon$
in gauge theory from the Hamiltonian formalism.
Section~\ref{sec:summary} is devoted to the conclusion and outlook.

\section{Quantum mechanics example}
\label{sec:qm}

In this section,
we describe the subtlety in
defining the real-time path integral
of the quantum mechanics on $S^1$.
This model has the subtlety of defining real-time path integral
that is similar to lattice  gauge theories \cite{Kanwar:2021tkd}.

\subsection{Subtlety of real-time path integral
  in quantum mechanics on $S^1$}
\label{sec:revi-subtl-defin}

We consider a one-dimensional quantum system with the action:
\begin{align}
  S[\phi] \equiv \frac{\beta}{2} \int_0^T dt\, (\partial_t \phi)^2,
  \label{eq:cont_action}
\end{align}
where $\phi(t)$ is the angular variable on $S^1$.
This model is equivalent to
the ordinary one-dimensional quantum mechanics in a periodic box
(see, e.g., \cite{Judge1963,Susskind:1964zz,Carruthers:1968my,Inomata1980,Ohnuki:1993nk,Tanimura:1993hf,
  Fukushima:2022zor})
by the identification
\begin{align}
  x(t) \equiv {L \over 2\pi} \phi(t),
\end{align}
where $L$ is the spatial extent of the system
and $\beta$ gives the particle mass $(2\pi)^2\beta/L^2$.
We here concentrate on the free case for simplicity.
The corresponding Hamiltonian of the system is
\begin{align}
  H \equiv \frac{1}{2\beta}p_{\phi}^2,
\end{align}
where $p_\phi$ is the conjugate momentum of $\phi$.
In quantum mechanics,
the plane waves $\{\exp(in\phi)\}_{n\in \mathbb{Z}}$
are the eigenfunctions of the momentum operator,
which in this case diagonalize the Hamiltonian
with the energy levels:
\begin{align}
  E_n \equiv \frac{1}{2\beta}n^2.
\end{align}

To define the path integral,
we discretize the time $T=N a$
and introduce the $U(1)$ variables
$U_\ell \equiv e^{i \phi_\ell}$, where
$\phi_\ell = \phi(a\ell)$ $(\ell=0,\cdots,N)$.
The transition amplitude from level $n_i$ to $n_f$:
\begin{align}
  A_{n_f,n_i}(T) &\equiv \langle n_f|e^{-i \hat{H} T}|n_i\rangle
                   \label{eq:amp}
\end{align}
may be expressed on the lattice naively as:
\begin{align}
  A^{\rm (lat)}_{n_f,n_i}(T) \equiv
  {\cal N} \int (dU)\,
  e^{i S(U)} (U_N^*)^{n_f}U_0^{n_i},
  \label{eq:A_lat}
\end{align}
where
\begin{align}
  (dU) &\equiv \prod_{\ell=0}^{N} dU_{\ell}
         \equiv \prod_{\ell=0}^{N} \frac{d\phi_\ell}{2\pi},\\
  S(U) &\equiv \frac{\beta}{2a} \sum_{\ell=0}^{N-1} |U_{\ell+1}-U_\ell|^2
         =-{\beta \over a} \sum_{\ell=0}^{N-1} {\rm Re}(U_{\ell+1}U_\ell^*) + {\rm const}.
         \label{eq:lat_action}
\end{align}
The normalization factor ${\cal N}$ can be determined
by demanding $A^{\rm (lat)}_{n_f,n_i}(0)=\delta_{n_f,n_i}$.

To obtain the analytic expression of $A^{\rm (lat)}_{n_f,n_i}(T)$,
we expand the exponential in terms of characters:
\begin{align}
  e^{-i(\beta/ a) {\rm Re}\, U}
  = \sum_{n\in \mathbb{Z}} I_n\Big({-i\beta\over a}\Big) U^n,
\end{align}
where $I_n(\beta)$ is the modified Bessel function of the first kind.
The integration in \eqref{eq:A_lat} can be performed analytically to give:
\begin{align}
  A^{\rm (lat)}_{n_f,n_i}(T)
  &= {\cal N} \delta_{n_f, n_i}
    I^N_{n_f}\Big({-i\beta\over a}\Big).
    \label{eq:analytical_qm}
\end{align}

The function $A^{\rm (lat)}_{n_f,n_i}(T)$ is an analytic function of the coupling $\beta$
for finite $a$;
however, it is not in the limit $a \rightarrow 0$.
This can be seen in the asymptotic expansion of $I_n(z)$
for $|z|\rightarrow \infty$ \cite{Inomata1979}:
\begin{align}
  I_n(z)
  &\sim
    \frac{e^z}{\sqrt{2\pi z}} \sum_{k\geq 0}
    \frac{\Gamma(n+k+1/2)}{k!\, \Gamma(n-k+1/2)}\Big(\frac{-1}{2z}\Big)^k\nonumber\\
  &~~~ \pm i e^{\pm i n \pi} \frac{e^{-z}}{\sqrt{2\pi z}} \sum_{k\geq 0}
    \frac{\Gamma(n+k+1/2)}{k!\, \Gamma(n-k+1/2)}\Big(\frac{1}{2z}\Big)^k.
    \label{eq:asymptotic2}
\end{align}
The plus signature applies for $-\pi/2<{\arg} z<3\pi/2$,
and the negative signature for $-3\pi/2<{\arg} z<\pi/2$.
For $|{\arg}\,z|<\pi/2$,
including the imaginary-time case (${\arg}\,z = 0$),
the second term will be completely
irrelevant because of the exponential factor.
However, at ${\arg}\,z=-\pi/2$, which is the case for eq.~\eqref{eq:analytical_qm},
the second term also contributes equally to the first term.
Therefore, the result will be different depending on
how we approach the real-time continuum limit.
To get the correct continuum limit,
one can modify the kinetic term \cite{Inomata1979, Junker1987}
by introducing a slight imaginary part
\begin{align}
  \beta \rightarrow e^{i\varepsilon} \beta \quad (\varepsilon>0). \label{eq:beta_ieps}
\end{align}
We first take the $a\rightarrow 0$ limit
keeping $\varepsilon$ finite,
and then take the $\varepsilon \rightarrow +0$ limit.
In fact, 
for $|{\arg} z|<\pi/2$,
\begin{align}
  I_n(z)/I_0(z)
  \sim 1-\frac{n^2}{2}{1\over z} + \cdots,
\end{align}
which in our case gives
\begin{align}
  \bigg[I_n\Big(\frac{-ie^{i\varepsilon}\beta}{a}\Big)
  /I_0\Big(\frac{-ie^{i\varepsilon}\beta}{a}\Big) \bigg]^N
  \sim \Big[1-i e^{-i\varepsilon}\frac{n^2}{2}{a\over\beta}  +\cdots\Big]^N
  \xrightarrow{a\rightarrow0} \exp\Big[-i e^{-i\varepsilon}\frac{n^2 t}{2 \beta}\Big].
  \label{eq:limit}
\end{align}
Therefore,
\begin{align}
  A^{\rm (lat)}_{n_f,n_i}(T)
  \xrightarrow{a\rightarrow0}
  \delta_{n_f, n_i}
  \exp \big[ -i e^{-i\varepsilon} E_{n_f} T \big]
  \xrightarrow{\varepsilon\rightarrow +0}
  \delta_{n_f, n_i}
  \exp \big[ -i E_{n_f} T \big],
  \label{eq:ampl_qm}
\end{align}
which is the desired real-time amplitude.

Note that we will not obtain the correct continuum amplitude
if we take $a\rightarrow 0$ exactly on $\varepsilon=0$ \cite{Kanwar:2021tkd}.
In this case, the amplitude $A^{\rm (lat)}_{n_f,n_i}(T)$ becomes
a singular function with a highly oscillatory behavior
because of the second term in eq.~\eqref{eq:asymptotic2}.

\subsection{
  $i\varepsilon$ in
  the derivation of the path integral}
\label{sec:explanation}

Although the argument in subsection~\ref{sec:revi-subtl-defin} is mathematically correct,
it may be uncertain why such $i\varepsilon$
becomes required to obtain the correct continuum theory
for the discretized action \eqref{eq:lat_action}.
In this subsection, we argue that,
starting from the Hamiltonian formalism,
we can understand the role of the $i\varepsilon$ as
manifestly suppressing the contributions from singular paths, 
for which
the discretized action
can give different values
from those of the actual continuum action.%
\footnote{
  The relation between path integral and
  the Hamiltonian formalism for a compact variable
  was argued in \cite{Inomata1979},
  but not was used to explain the meaning of the $i\varepsilon$.
}

We consider the Feynman kernel
for an infinitesimal time increment $a$:
\begin{align}
  \langle U' | e^{-i a \hat{H}} | U \rangle,
\end{align}
where $| U \rangle$ is the eigenstate of the unitary operator
$\hat{U}$, $\hat{U} | U \rangle = U | U \rangle$,
that satisfies the commutation relation:
\begin{align}
  [\hat{U}, \hat{p}_\phi] = \hat{U}.
\end{align}
By inserting the momentum eigenstates $|n \rangle$:
\begin{align}
  \langle U | n \rangle \equiv U^n \quad (n\in \mathbb{Z}),
\end{align}
we have
\begin{align}
  \langle U' | e^{-i a \hat{H}} | U \rangle
  &= \sum_{n\in\mathbb{Z}} \exp\Big[ \frac{-ia}{2\beta}n^2
  + i n (\phi' - \phi) \Big] \nonumber\\
  &=e^{i\pi/4} \sqrt{\frac{2\pi \beta}{-a }}
  \sum_{w\in\mathbb{Z}}
  \exp \Big[ i \frac{\beta (\phi' - \phi +2\pi w )^2}{2a}\Big],
  \label{eq:summation_mid}
\end{align}
where we write $U=\exp(i\phi)$, $U'=\exp(i\phi')$ with $\phi,\phi'\in[-\pi,\pi)$
and have used the Poisson summation formula to obtain the second line.
Although the kernel \eqref{eq:summation_mid} is not well-defined as an
ordinary function because the theta function
\begin{align}
  \vartheta(v,\tau) \equiv \sum_{n\in\mathbb{Z}} e^{\pi i n^2 \tau } e^{2\pi i n v}
  \label{eq:theta}
\end{align}
is only analytic for ${\rm Im}\,\tau>0$,
the kernel has a definite meaning as a distribution.
To see this,
it should be sufficient to check the Fourier integral
in which the kernel is multiplied by the plane waves
because all the state vectors can be expressed as
a linear combination of these basis vectors.
For the kernel \eqref{eq:summation_mid},
we trivially obtain
\begin{align}
  \int dU'\, ({U'}^{*})^n \langle U' | e^{-i a \hat{H}} | U \rangle
  = (U^{*})^n e^{-\frac{ia}{2\beta}n^2} ,
  \label{eq:fourier_orig}
\end{align}
which is a well-defined number for given $n$ and $U$,
and thus establishes the definite meaning of the kernel
as a distribution.

We can now understand the need of $i\varepsilon$
discussed in section~\ref{sec:revi-subtl-defin}
with distributional terms. 
In fact, the naive real-time path integral in section~\ref{sec:revi-subtl-defin}
amounts to replacing the kernel \eqref{eq:summation_mid}
by the expression:
\begin{align}
  \langle U' | e^{-i a \hat{H}} | U \rangle \to
  e^{i\pi/4} \sqrt{\frac{2\pi \beta}{-a }}
  \exp\Big[\frac{i\beta}{a}\big[1- {\rm Re}\,(U {U'}^*)\big]\Big].
\end{align}
This replacement cannot be justified as
a distributional relation because, 
as in section~\ref{sec:revi-subtl-defin},
the Fourier integral gives
\begin{align}
  &\int dU'\, ({U'}^{*})^n
    e^{i\pi/4} \sqrt{\frac{2\pi \beta}{-a }}
  \exp\Big[\frac{i\beta}{a}\big[1- {\rm Re}\,(U {U'}^*)\big]\Big] \nonumber\\
  &=
  e^{i\pi/4} \sqrt{\frac{2\pi \beta}{-a }} e^{-in\phi}
  e^{i\beta/a} I_n\Big(\frac{-i\beta}{a}\Big) \nonumber\\
  &\sim e^{-in\phi} \big[
    e^{\frac{-ia}{2\beta}(n^2-\frac{1}{4})}
    -i(-1)^n e^{\frac{2i\beta}{a}}e^{\frac{ia}{2\beta}(n^2-\frac{1}{4})}
    \big],
  \label{eq:fourier_replaced_noeps}
\end{align}
which has the $n$ and $\phi$ dependent second term.
However, the first term has the correct $n$ and $\phi$ dependence,
and the second term can be removed by the $i\varepsilon$.
We thus have the distributional identity
after correcting the shift of the zero-point energy:%
\footnote{
  The zero-point energy was absorbed in the
  normalization factor ${\cal N}$ in section~\ref{sec:revi-subtl-defin}.
}
\begin{align}
  \langle U' | e^{-i a \hat{H}} | U \rangle
  =
  \lim_{\varepsilon \rightarrow +0}
  e^{\frac{-ia}{8e^{i\varepsilon \beta}}}
  e^{i\pi/4} \sqrt{\frac{2\pi e^{i\varepsilon}\beta}{-a }}
  \exp\Big[\frac{ie^{i\varepsilon}\beta}{a}\big[1- {\rm Re}\,(U {U'}^*)\big]\Big].
  \label{eq:eq_distribution}
\end{align}
This justifies the use of the discretized action \eqref{eq:lat_action}
under the $i\varepsilon$.
The rest of this section is devoted to systematically deriving
this distributional equality.

We begin with introducing the $i\varepsilon$
and regarding the original kernel \eqref{eq:summation_mid}
as the $\varepsilon \to +0$ limit:
\begin{align}
  \langle U' | e^{-i a \hat{H}} | U \rangle
  =
  \lim_{\varepsilon \rightarrow +0}
  \langle U' | e^{-i a \hat{H}} | U \rangle \big|_{\beta \to e^{i\varepsilon}\beta}.
  \label{eq:eps_zero_lim}
\end{align}
The expression inside the limit now becomes a well-defined function,
and has a sharp peak around $U=U'$
for an infinitesimal $a$.
This allows us to rewrite the expression as
\begin{align}
    \langle U' | e^{-i a \hat{H}} | U \rangle |_{\beta \to e^{i\varepsilon}\beta}
  \approx
    e^{i\pi/4} \sqrt{\frac{2\pi e^{i \varepsilon} \beta}{-a }}
    \exp
    \Big( \frac{i e^{i \varepsilon} \beta}{2a}
    \lfloor \phi'-\phi \rfloor^2 \Big),
    \label{eq:phi_phip}
\end{align}
where the function $\lfloor \cdot \rfloor$ returns the value
in $[-\pi, \pi)$ modulo $2\pi$.
Relation \eqref{eq:phi_phip}
becomes a distributional equality for an infinitesimal $a$
because the contributions with nontrivial winding are
exponentially suppressed thanks to $\varepsilon>0$.

On the other hand, the kernel
\begin{align}
  \exp\Big[\frac{ie^{i\varepsilon}\beta}{a}\big[1- {\rm Re}\,(U {U'}^*)\big]\Big]
  =
  \exp\Big[\frac{ie^{i\varepsilon}\beta}{a}\big[1- \cos(\phi-\phi')\big]\Big]
  \label{eq:cos_kernel}
\end{align}
has a similar functional dependence to eq.~\eqref{eq:phi_phip};
the function \eqref{eq:cos_kernel} has a sharp peak around
$U=U'$ for an infinitesimal $a$,
which allows us to expand the cosine in powers of $\lfloor\phi-\phi'\rfloor$
and convert the Fourier integral to a Gaussian integral:
\begin{align}
  &\int_{-\pi}^\pi \frac{d\phi'}{2\pi}\,
    e^{-in\phi'}
    \exp\Big[\frac{ie^{i\varepsilon}\beta}{a}\big[1- \cos(\phi-\phi')\big]\Big]\nonumber\\
  &=\int_{-\pi}^\pi \frac{d\phi'}{2\pi}\,
    e^{-in\phi'}
    \exp\Big[
    \frac{ie^{i\varepsilon}\beta}{2a} \lfloor\phi-\phi'\rfloor^2
    -\frac{ie^{i\varepsilon}\beta}{24a} \lfloor\phi-\phi'\rfloor^4 + \cdots
    \Big] \nonumber\\
  &\approx
    e^{-in\phi}e^{-\frac{ia}{2\beta}n^2}
    \int_{-\infty}^\infty \frac{d\phi''}{2\pi}\,
    \exp\Big[
    \frac{ie^{i\varepsilon}\beta}{2a} {\phi''}^2
    \Big]
    \Big(
    1-\frac{ie^{i\varepsilon}\beta}{24a} {\phi''}^4 + \cdots
    \Big).
    \label{eq:overall_factor}
\end{align}
We see the desired $n$ and $\phi$ dependence in front of the Gaussian integral.
The remaining integral only gives an overall constant
that includes the shift of the zero-point energy:
\begin{align}
  \int_{-\infty}^\infty \frac{d\phi''}{2\pi}\,
  \exp\Big[
  \frac{ie^{i\varepsilon}\beta}{2a} {\phi''}^2
  \Big]
  \Big(
  1-\frac{ie^{i\varepsilon}\beta}{24a} {\phi''}^4 + \cdots
  \Big)
  &=
  \Big(
  1+\frac{ia}{8e^{i\varepsilon}\beta}+\cdots
  \Big)
  \sqrt{\frac{-a }{2\pi ie^{i\varepsilon}\beta}} \nonumber\\
  &\sim
  \exp\Big(
  \frac{ia}{8e^{i\varepsilon}\beta}
  \Big)
  \sqrt{\frac{-a }{2\pi ie^{i\varepsilon}\beta}}.
  \label{eq:nonlinearlity}
\end{align}
Correcting this constant gives the
distributional relation \eqref{eq:eq_distribution}.

Note the ordering of the limit.
The distributional relation \eqref{eq:eq_distribution}
is for an infinitesimal $a$ and for $\varepsilon>0$,
and thus we first take the $a \rightarrow 0$ limit
keeping $\varepsilon > 0$.
Correspondingly,
we take the $\varepsilon \to +0$ outside the path integral
once we adopt the discretized action \eqref{eq:lat_action}:
\begin{align}
  A_{n_f,n_i}(T)
  &= {\cal N}
    \lim_{\varepsilon\rightarrow +0}
    \lim_{a \to +0}
    \int(dU)\,
    e^{-i(e^{i \varepsilon}\beta /a) \sum_{\ell=0}^{N-1} {\rm Re}(U_{\ell+1}U_\ell^*)}
    (U_N^*)^{ n_f }(U_0)^{ n_0 }.
    \label{eq:derived_path_integral}
\end{align}
This establishes the necessity of $i\varepsilon$
in the real-time path integral discussed in section~\ref{sec:revi-subtl-defin}.

From the above derivation,
we can understand the role of $i\varepsilon$
for the discretized action \eqref{eq:summation_mid} as follows.
Firstly, as expected, large fluctuations
basically do not contribute to the amplitude
in the original theory,
which can be seen from the facts that
the kernel \eqref{eq:summation_mid} becomes
the periodic delta function at $a=0$
and that we are able to safely introduce
the $i\varepsilon$ in eq.~\eqref{eq:eps_zero_lim}.
On the other hand, the discretized action \eqref{eq:lat_action}
is designed in such a way that
it reproduces the continuum action for smooth fields
but not necessarily for these large fluctuations.
As we have discussed, this difference in fact 
changes the distributional property of the kernel,
and we thus need to suppress
the contributions from singular paths in advance
with the $i\varepsilon$ when using the action \eqref{eq:lat_action}.
As shown in eq.~\eqref{eq:overall_factor},
the nonlinearity of the cosine function
only affects the overall constant.

\section{Gauge theory case}
\label{sec:field_theory}

In this section, we consider the gauge theory.
The structure is basically the same as in the quantum mechanical system
discussed in section~\ref{sec:qm}.

\subsection{Necessity of the $i\varepsilon$ in lattice gauge theories}
\label{sec:i_eps_lattice}

The lattice Yang-Mills action for $SU(N_c)$ gauge group
in four-dimensional Minkowski spacetime can be given by
\cite{Wilson:1974sk, Berges:2006xc}:
\begin{align}
  S(U) &\equiv
         \beta_t \sum_x \sum_i \big[
         1-\frac{1}{N_c}{\rm Re}\,{\rm tr}\,[U_{x,i}U_{x+i,t}U_{x+t,i}^\dagger U_{x,t}^\dagger]
         \big] \nonumber\\
       &~~~ - \beta_s \sum_x \sum_{i<j} \big[
         1-\frac{1}{N_c}{\rm Re}\,{\rm tr}\,[U_{x,i}U_{x+i,j}U_{x+j,i}^\dagger U_{x,j}^\dagger]
         \big],
\end{align}
where
\begin{align}
  \beta_t \equiv \frac{a}{a_0}\frac{2N_c}{g^2},\\
  \beta_s \equiv \frac{a_0}{a}\frac{2N_c}{g^2}
\end{align}
with the spatial lattice spacing $a$ and the time increment $a_0$.
We take the normalization of the generators as ${\rm tr}\,T^aT^b = (1/2)\delta^{ab}$.
The local Boltzmann factor
can be expanded with the characters $\chi_R$ as:
\begin{align}
  e^{i (-1)^r  (\beta_r/N_c) {\rm Re}\,{\rm tr}\,U}
  = \sum_{R:{\rm irrep}} d_R \,c_R(i(-1)^r \beta_r) \chi_R(U),
\end{align}
where $r=t,s$ labels the timelike and spacelike directions:
$(-1)^t=-1,(-1)^s=+1$
and $d_R$ is the dimension of the irreducible representation $R$.
The functions $c_R$ are given by
\cite{Brower:1981vt, Drouffe:1983fv}:
\begin{align}
  c_R( i(-1)^r \beta_r) =
  \frac{1}{d_R}\sum_{n \in \mathbb{Z}} \det_{1\leq j,k\leq N_c} I_{\ell_k-k+j+n}(i (-1)^r  \beta_r/N_c),
  \label{eq:charac_coeff}
\end{align}
where $\ell_k$ $(\ell_1\geq \ell_2\geq\cdots \ell_{N_c-1}\geq \ell_{N_c} \equiv 0)$
is the number of boxes in the $k$-th row of the Young diagram
representing the irreducible representation $R$ of $SU(N_c)$.
Since
$\beta_r\rightarrow \infty$
in the continuum limit of asymptotically free theories,
we again confront the subtlety
coming from the asymptotic expansion of the modified Bessel function.
To obtain the continuum limit,
we introduce slight imaginary parts:
\begin{align}
  \beta_t &\rightarrow e^{i\varepsilon} \beta_t, \label{eq:eps_gauge_t}\\
  \beta_s &\rightarrow e^{-i\varepsilon} \beta_s.\label{eq:eps_gauge_s}
\end{align}
It is noteworthy that
we should give the infinitesimal imaginary part
also for the spacelike plaquettes.%
\footnote{
  This point was not mentioned in \cite{Kanwar:2021tkd}.
}
The sign of the imaginary part for the timelike plaquettes
can be justified by the argument in subsection~\ref{sec:operator}.
To explain the sign for the spacelike plaquettes,
one can use the symmetry argument
that, since the continuum theory should be
Lorentz invariant,
the asymptotic formula should be the same
for the timelike and spatial plaquettes.
The signs agree with those given by the ordinary $i\varepsilon$ in the continuum theory.

\subsection{Convergence properties of the Wilson action}
\label{sec:convergence}

To confirm the convergence properties related to the $i\varepsilon$,
we consider the $SU(N_c)$ Wilson theory in two-dimensional spacetime with $N_c=2,3$.
We only have the timelike plaquettes in this case,
and we set 
\begin{align}
  \beta_t = \frac{2 N_c}{(a g)^2},
\end{align}
treating spacetime uniformly.

The expectation value of the $\ell \times \tau$ Wilson loop 
with the physical area
$A \equiv \ell \tau a^2$, $W_A$, 
can be expressed by
the characters of the trivial and fundamental representations
\cite{Gross:1980he, Drouffe:1983fv}:
\begin{align}
  \langle W_A \rangle
  = N_c \Big(
  \frac{c_{\rm fund}(-i e^{i\varepsilon}\beta_t)}{c_{\rm triv}(-i e^{i\varepsilon}\beta_t)}
  \Big)^{\ell \tau},
  \label{eq:formula}
\end{align}
for which the continuum limit is known from the analysis of
the heat-kernel action \cite{Menotti:1981ry,Kanwar:2021tkd}:
\begin{align}
  \lim_{\varepsilon \rightarrow +0} \lim_{a \rightarrow 0}  \langle W_A \rangle
  =
  N_c e^{-i (N/4)(1-1/N^2) \, g^2 A}.
  \label{eq:cont_string} 
\end{align}
Since $g$ is dimensionful,
we fix $g=1$ in the following.

\begin{figure}[htb]
  \centering
  \includegraphics[width=80mm]{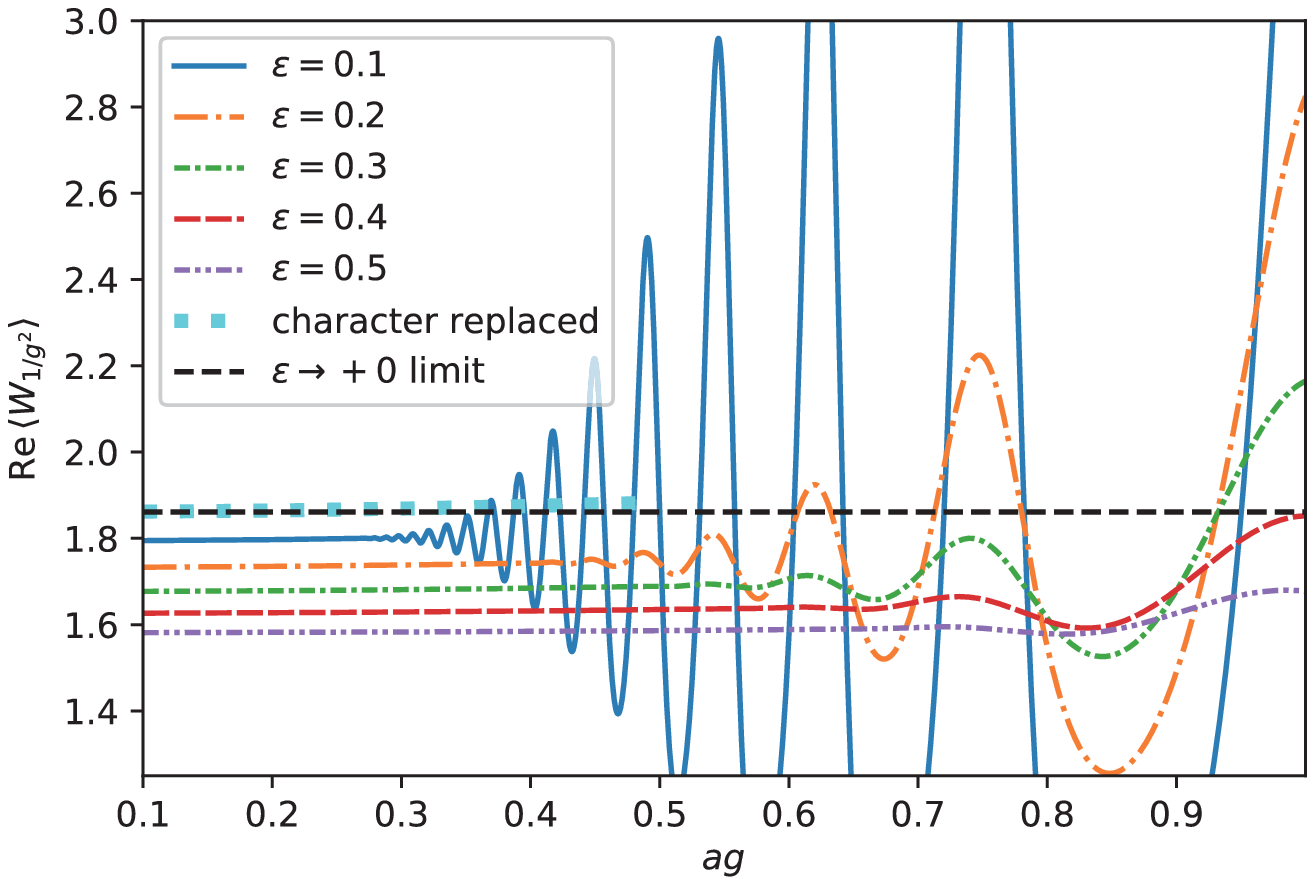}
  \vspace{5mm}
  \includegraphics[width=80mm]{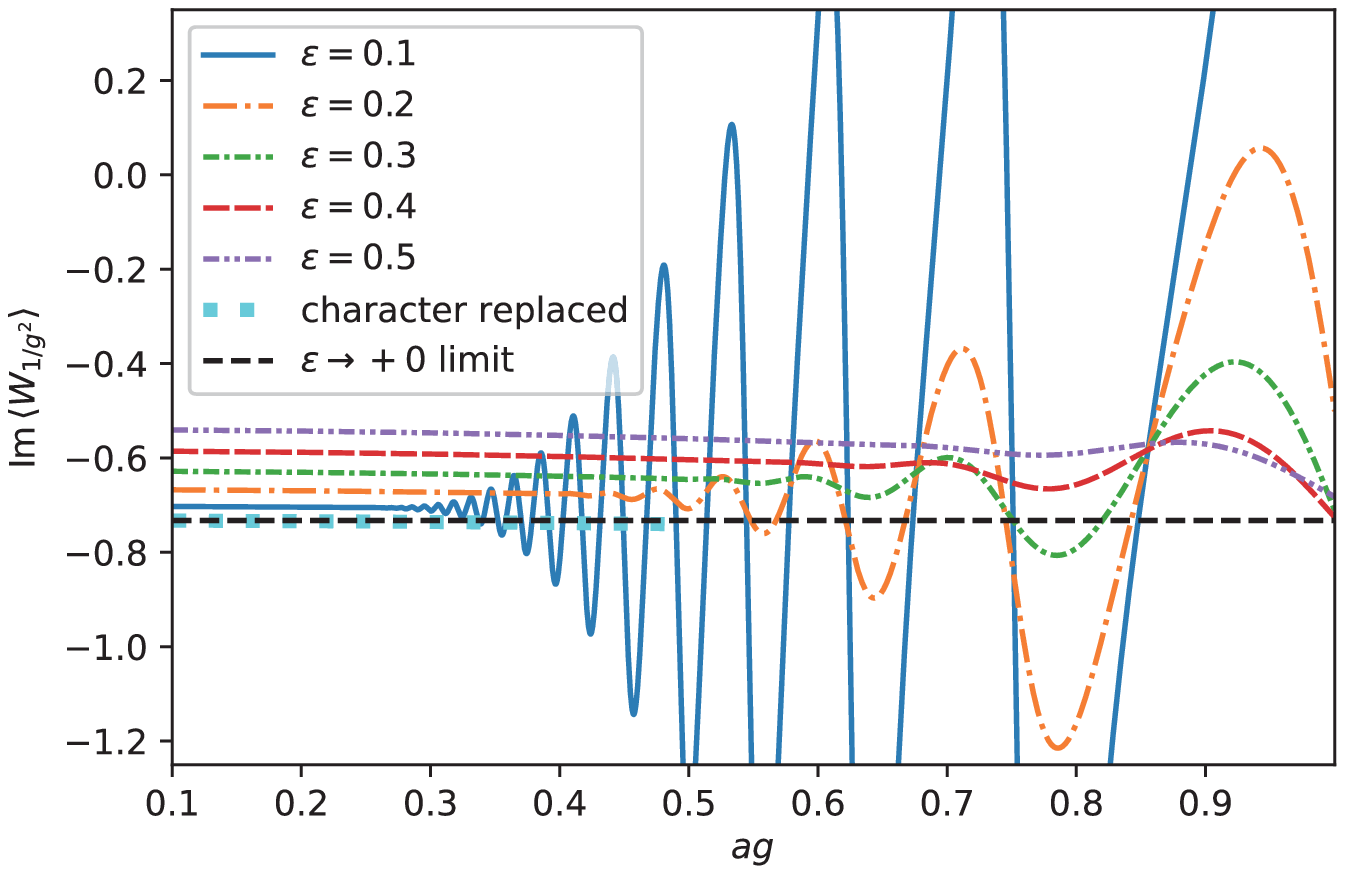} 
  \caption{
    \label{fig:re_w_su2}
    The expectation value of the Wilson loop $\langle W_{A} \rangle$
    with the area $A=1/g^2$
    evaluated with
    the analytic formula \eqref{eq:formula} for $SU(2)$.
    The values of $\varepsilon$ are varied to $\varepsilon = 0.1, \cdots, 0.5$.
    The cyan dotted line shows the $\varepsilon=0$ values
    with the modified Bessel function replaced by
    the asymptotic expansion dropping the unwanted part,
    which is drawn in the region where the asymptotic expansion
    gives a sufficient convergence up to the machine precision.
    The black dashed line shows the
    $a\rightarrow 0$, $\varepsilon \rightarrow +0$ value,
    eq.~\eqref{eq:cont_string}.
  }
\end{figure}

\begin{figure}[htb]
  \centering
  \includegraphics[width=80mm]{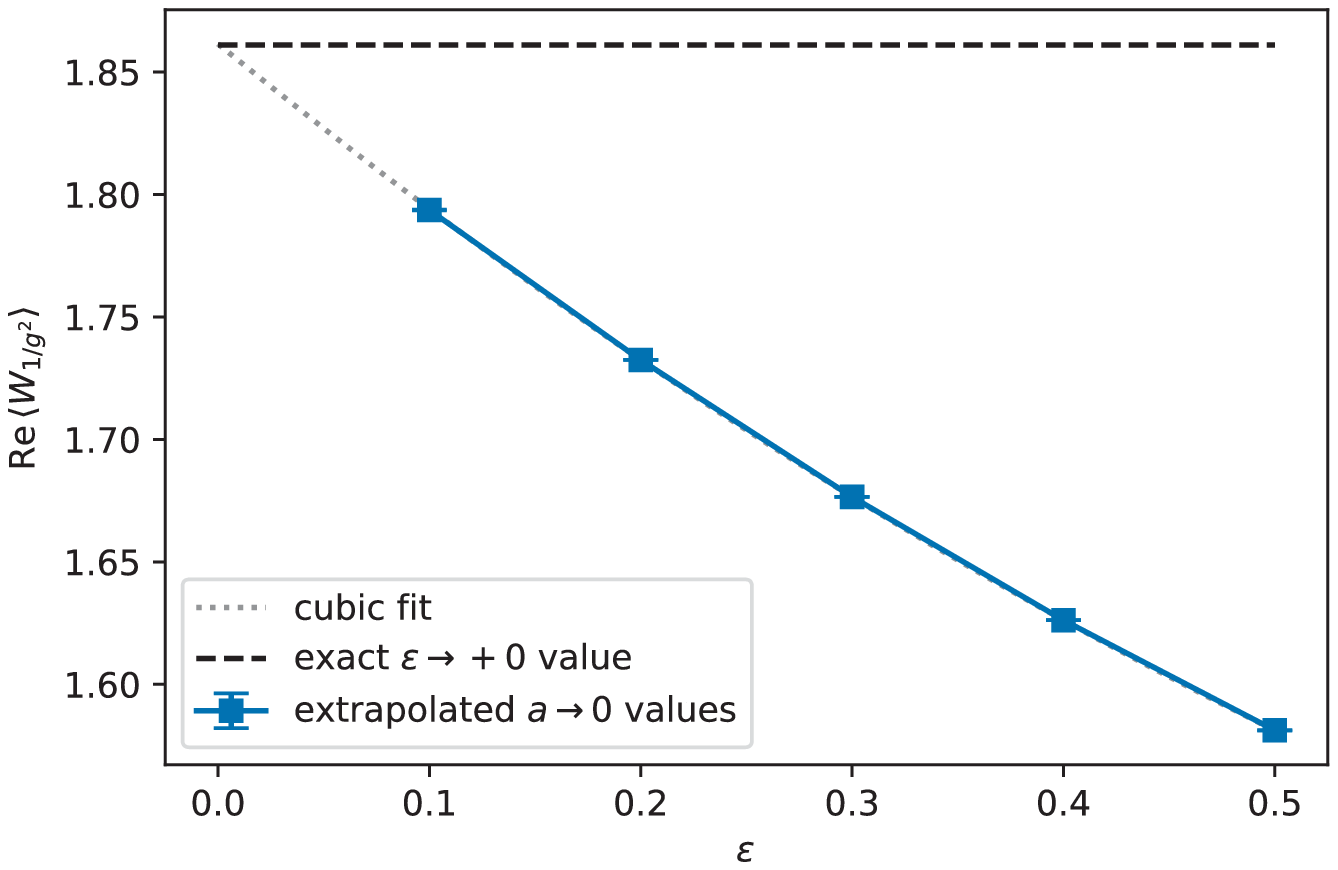}
  \vspace{5mm}
  \includegraphics[width=80mm]{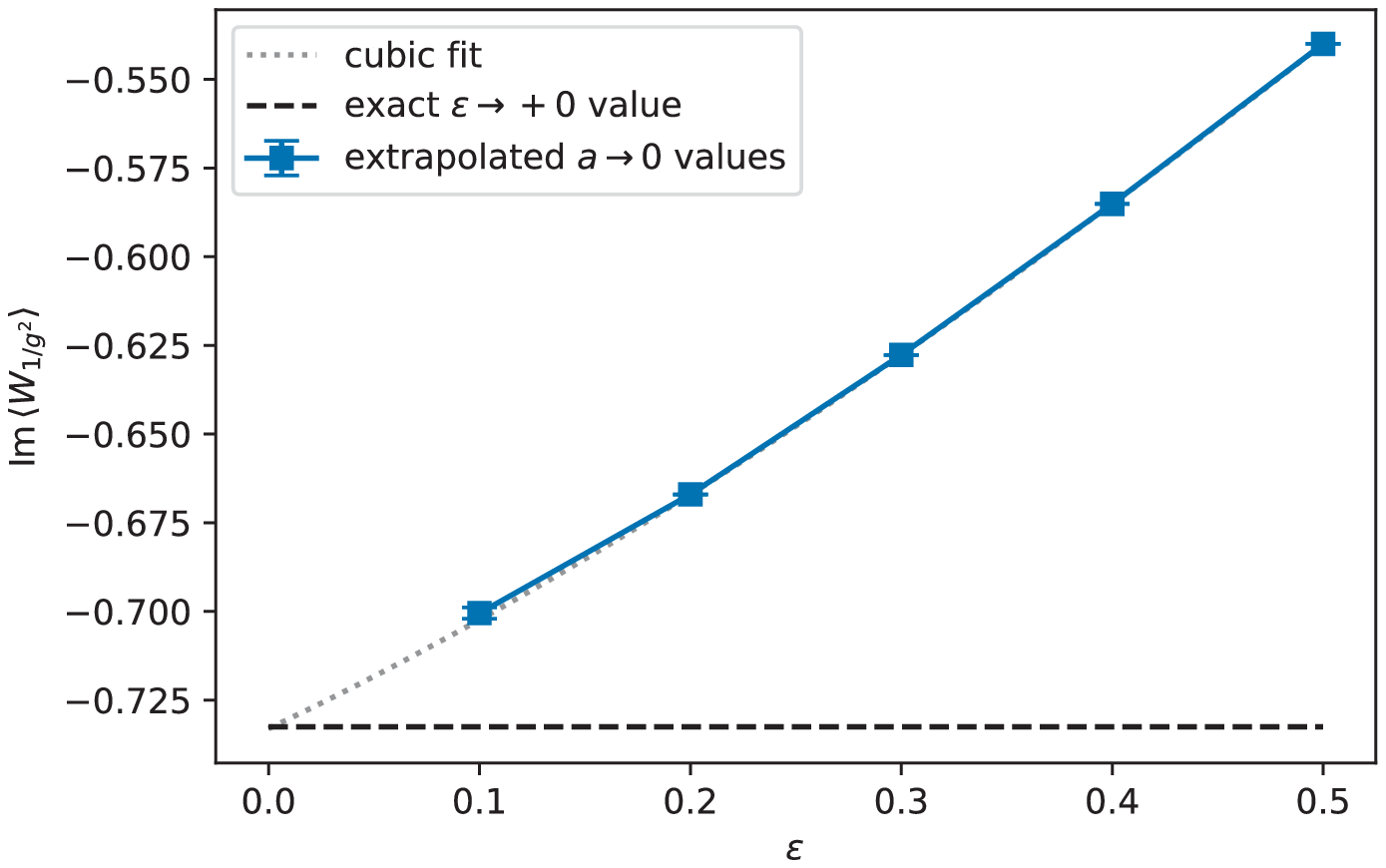} 
  \caption{
    The extrapolated $a\rightarrow0$ values of $\langle W_{1/g^2} \rangle$
    with various $\varepsilon$ for $SU(2)$.
    The $a\rightarrow0$ values are then fitted
    to obtain the final $\varepsilon\rightarrow +0$ result.
    The exact $\varepsilon \rightarrow +0$ value \eqref{eq:cont_string}
    is shown with the black dashed line for comparison.
    \label{fig:eps_extrapolation_su2}
  }
\end{figure}

\begin{figure}[htb]
  \centering
  \includegraphics[width=80mm]{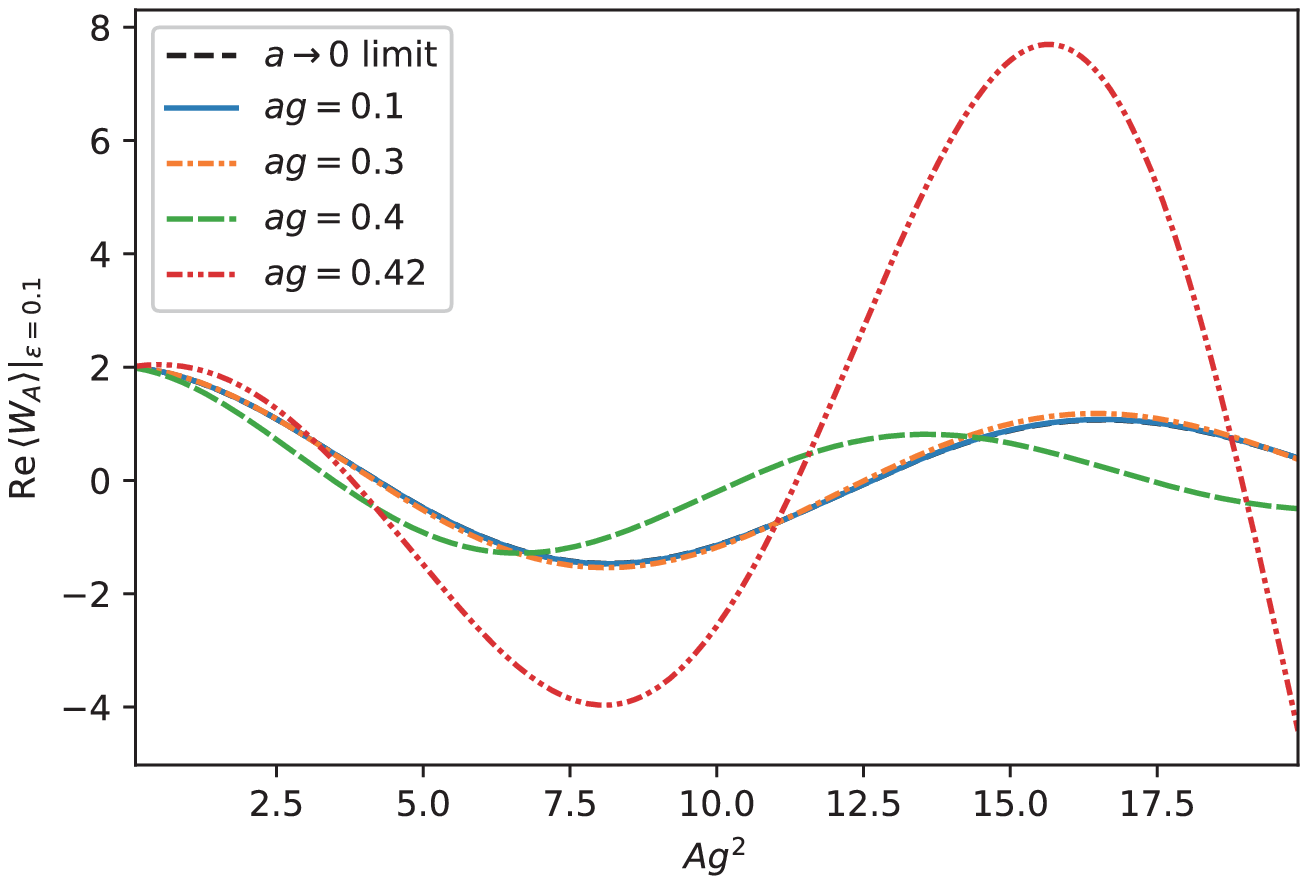}
  \vspace{5mm}
  \includegraphics[width=80mm]{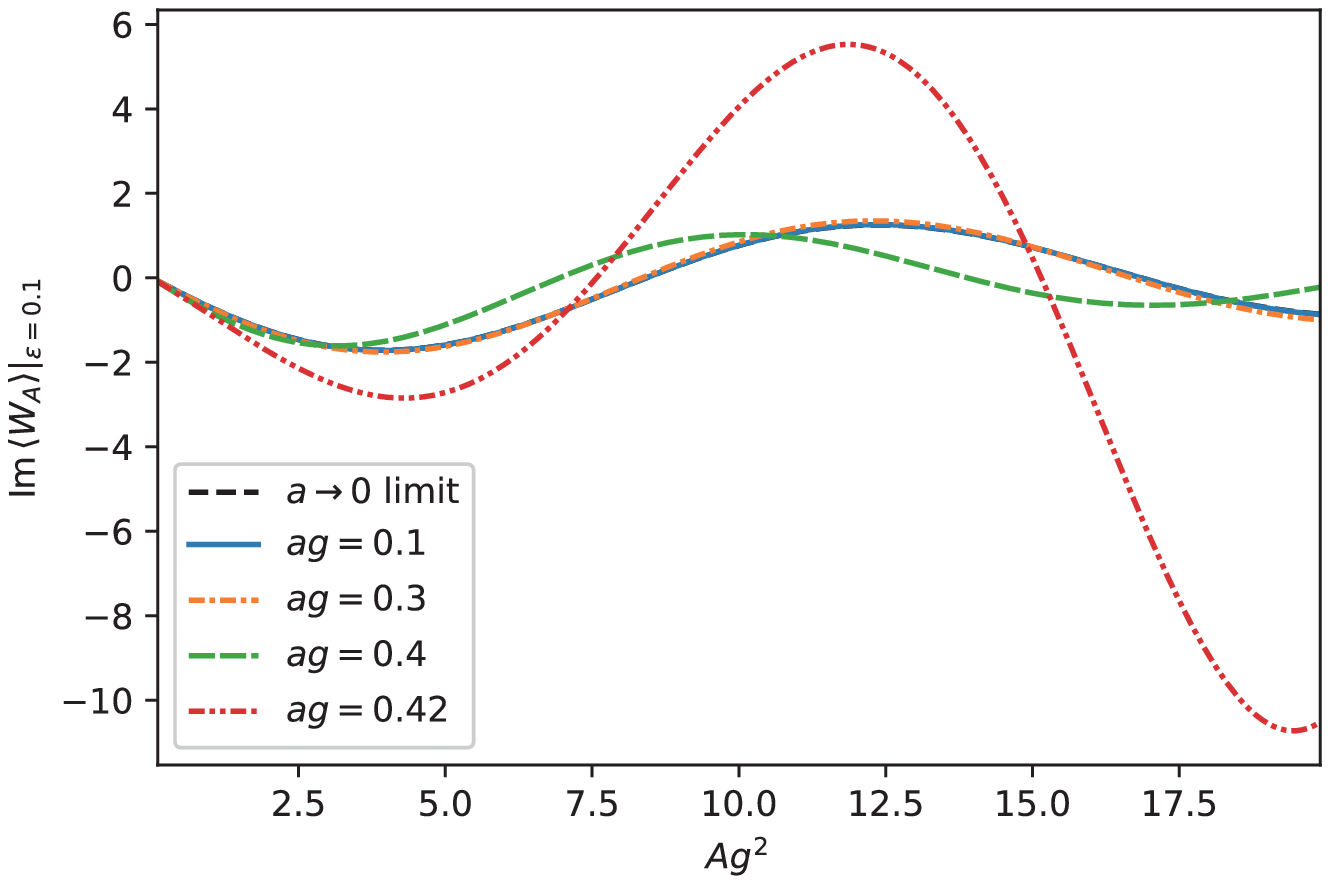}
  \caption{
    The area $A$ dependence of $\langle W_{A} \rangle$
    evaluated with various $a$ keeping $\varepsilon=0.1$ fixed ($N_c = 2$).
    \label{fig:eps01_su2}
  }
\end{figure}

\begin{figure}[htb]
  \centering
  \includegraphics[width=80mm]{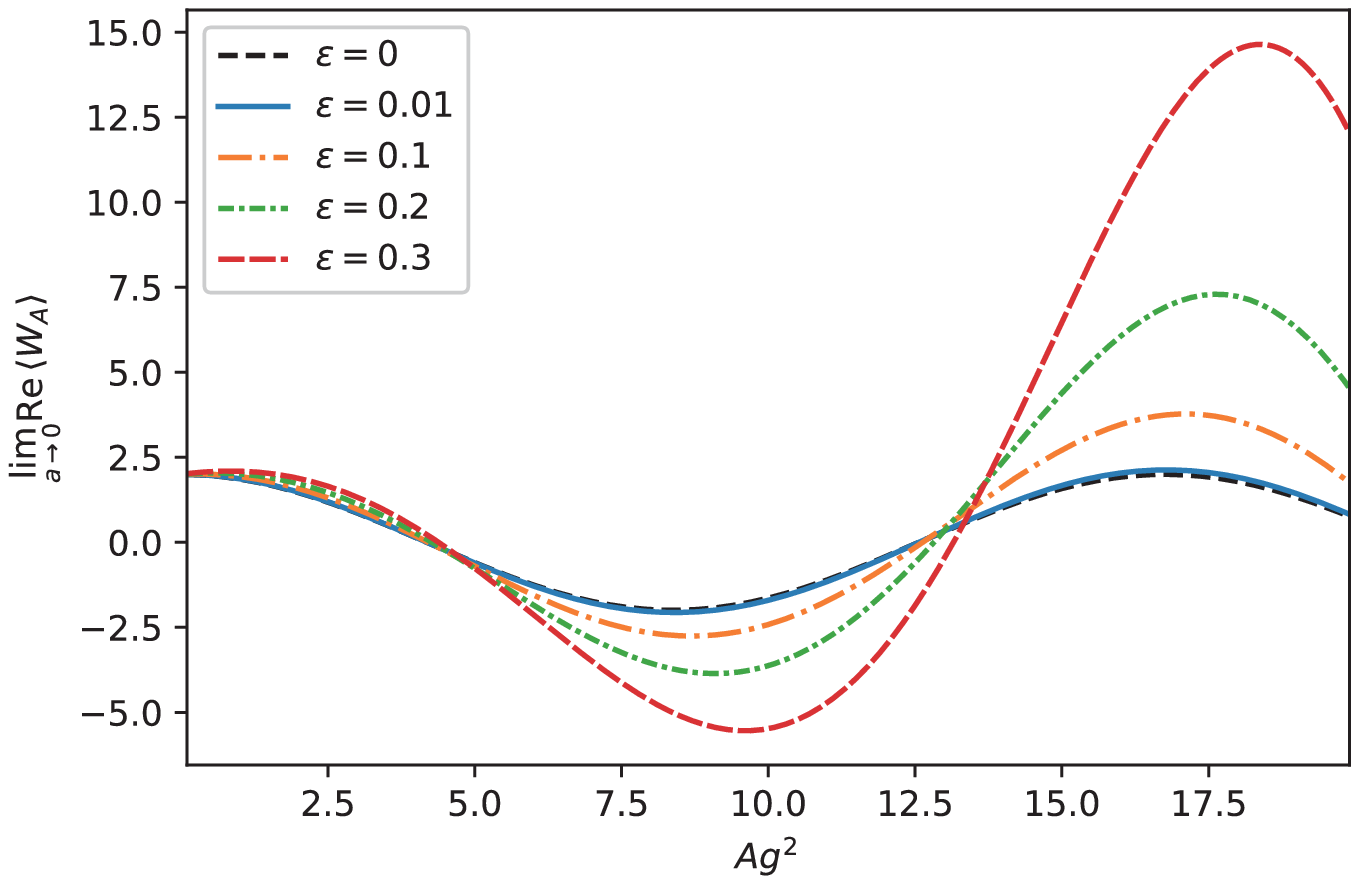}
  \vspace{5mm}
  \includegraphics[width=80mm]{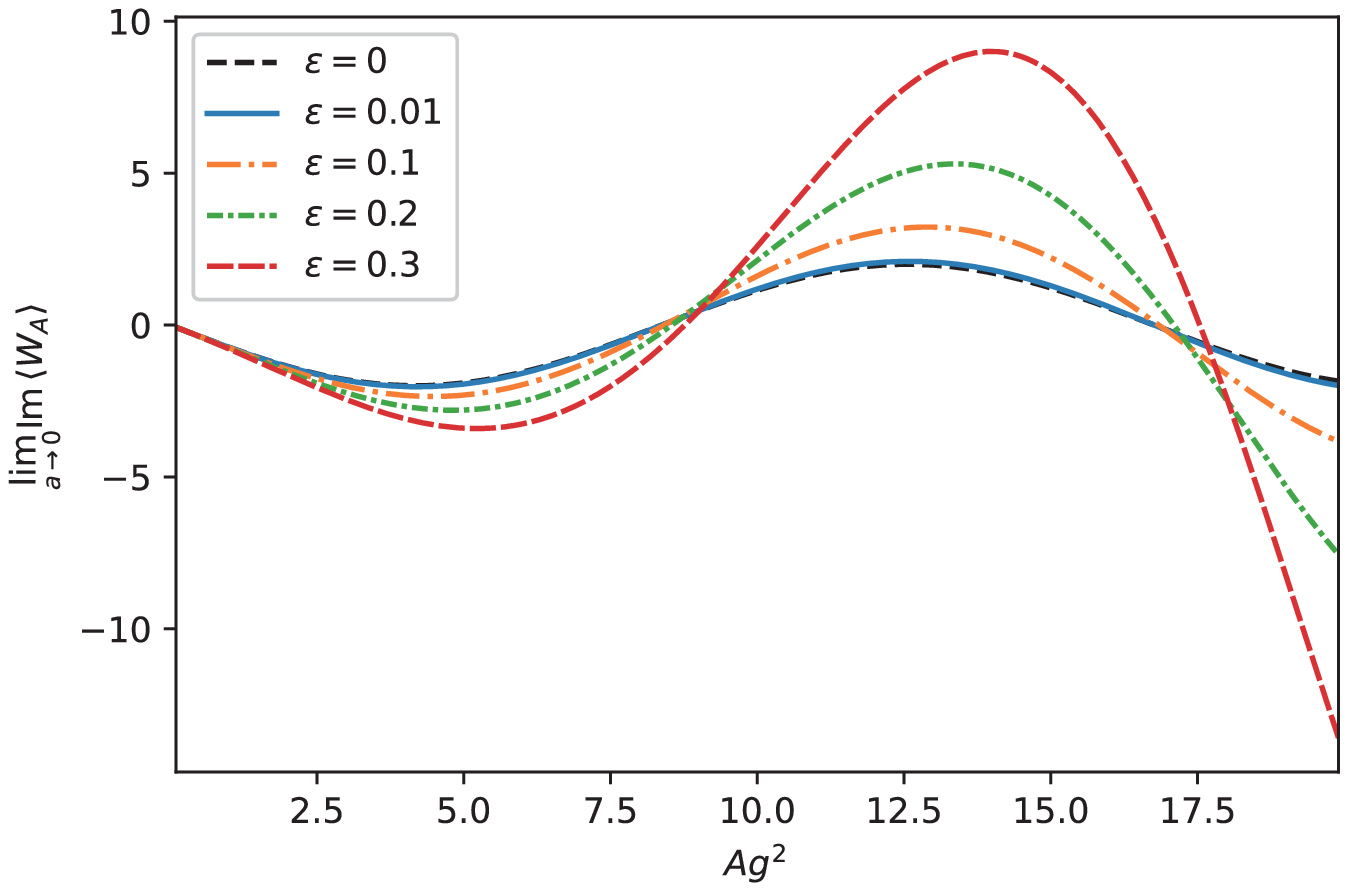}
  \caption{
    The area $A$ dependence of $\lim_{a\rightarrow 0} \langle W_{A} \rangle$
    evaluated with various $\varepsilon$ ($N_c = 2$).
    \label{fig:cont_su2}
  }
\end{figure}

We begin with $SU(2)$.
The character expansion coefficients \eqref{eq:charac_coeff}
has the well-known form for the spin-$j$ representation ($d_j = 2j+1$):
\begin{align}
  c_j( -i e^{i\varepsilon} \beta_t )
  = \frac{2 I_{2j+1}(-i e^{i\varepsilon} \beta_t)}{-i e^{i\varepsilon} \beta_t},
\end{align}
with which we can confirm the $a\rightarrow 0, \varepsilon \rightarrow +0$ limit
\eqref{eq:cont_string} from eq.~\eqref{eq:formula}:
\begin{align}
  \langle W_A \rangle
  \sim
  2 \Big( 1-\frac{3}{2}\frac{i e^{-i\varepsilon}a^2 g^2}{4} \Big)^{A/a^2}
  \xrightarrow{a \rightarrow 0, \varepsilon \rightarrow +0}
  2 e^{-i(3/8)g^2 A}.
  \label{eq:cont_string_su2}
\end{align}
Figure~\ref{fig:re_w_su2} shows the
expectation value $\langle W_{A} \rangle$ with the area $A=1$,
where the results are calculated
directly using the modified Bessel function for
various $\varepsilon$.
We see that, for relatively large $a$,
the unwanted part of the asymptotic expansion \eqref{eq:asymptotic2}
contributes to give oscillatory behavior.
This shows that in practice, for a given $a$,
we need to prepare $\varepsilon$ large enough so that
the unwanted part can be neglected.
On the other hand, instead of implementing the $i\varepsilon$,
we can expand the action in terms of the characters
and replace the modified Bessel function
with its asymptotic expansion
dropping the unwanted part in advance.
The corresponding result with $\varepsilon=0$ is shown with
the cyan dotted line in figure \ref{fig:re_w_su2}
for the region where the asymptotic expansion
gives sufficient convergence up to the machine precision.
The continuum value \eqref{eq:cont_string} is
shown with the black dashed line for comparison.
\noindent
For completeness, we perform the $\varepsilon \rightarrow +0$ extrapolation
of the $a \rightarrow 0$ limits.
To obtain the $a\rightarrow 0$ values for each $\varepsilon$,
we fit five points
$a = 0.1, 0.15, \cdots, 0.3$
with the linear function of $a^2$.
The systematic error is calculated from the
estimated variance of the fitting parameter.
The obtained values for the $A=1$ case 
are shown in figure~\ref{fig:eps_extrapolation_su2}.
\noindent
We fit these values with a quadratic and cubic functions of $\varepsilon$
to give the final $\varepsilon \rightarrow +0$ value.
We use the cubic result for the central value,
and take the difference from the quadratic value
as the estimate of the systematic error.
The chi-squared for the cubic fits are $\chi^2/{\rm DOF} = 3.3$ and $1.4$,
respectively, for the real and imaginary parts.
The obtained estimate
$\lim_{a\rightarrow 0, \varepsilon\rightarrow +0} \langle W_{A=1} \rangle \approx 1.86146(93) -0.7331(36) i$
agrees with the analytical value
$\lim_{a\rightarrow 0, \varepsilon\rightarrow +0} \langle W_{A=1} \rangle = 1.8610 - 0.7325 i$ within the estimated systematic error.
To see how the finite $a$ or $\varepsilon$
effect depends on $A$,
we also plot $\langle W_{A} \rangle$
with
various $a$ for $\varepsilon=0.1$ (figure~\ref{fig:eps01_su2})
and the $a\rightarrow 0$ values with various $\varepsilon$
(figure~\ref{fig:cont_su2}).
\noindent
\noindent
We see that the effect of finite $a$ or $\varepsilon$
becomes larger as we increase $A$.

\begin{figure}[htb]
  \centering
  \includegraphics[width=80mm]{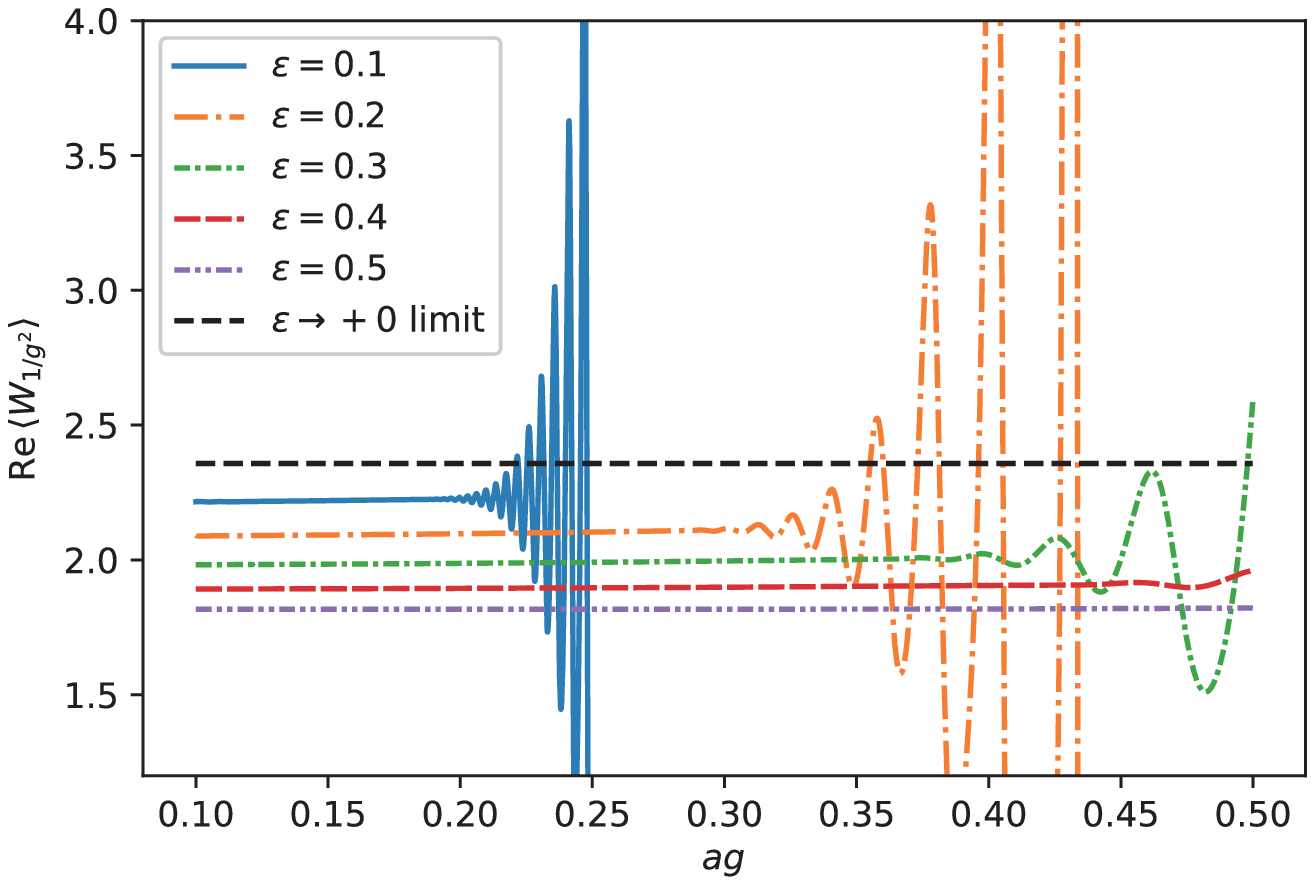}
  \vspace{5mm}
  \includegraphics[width=80mm]{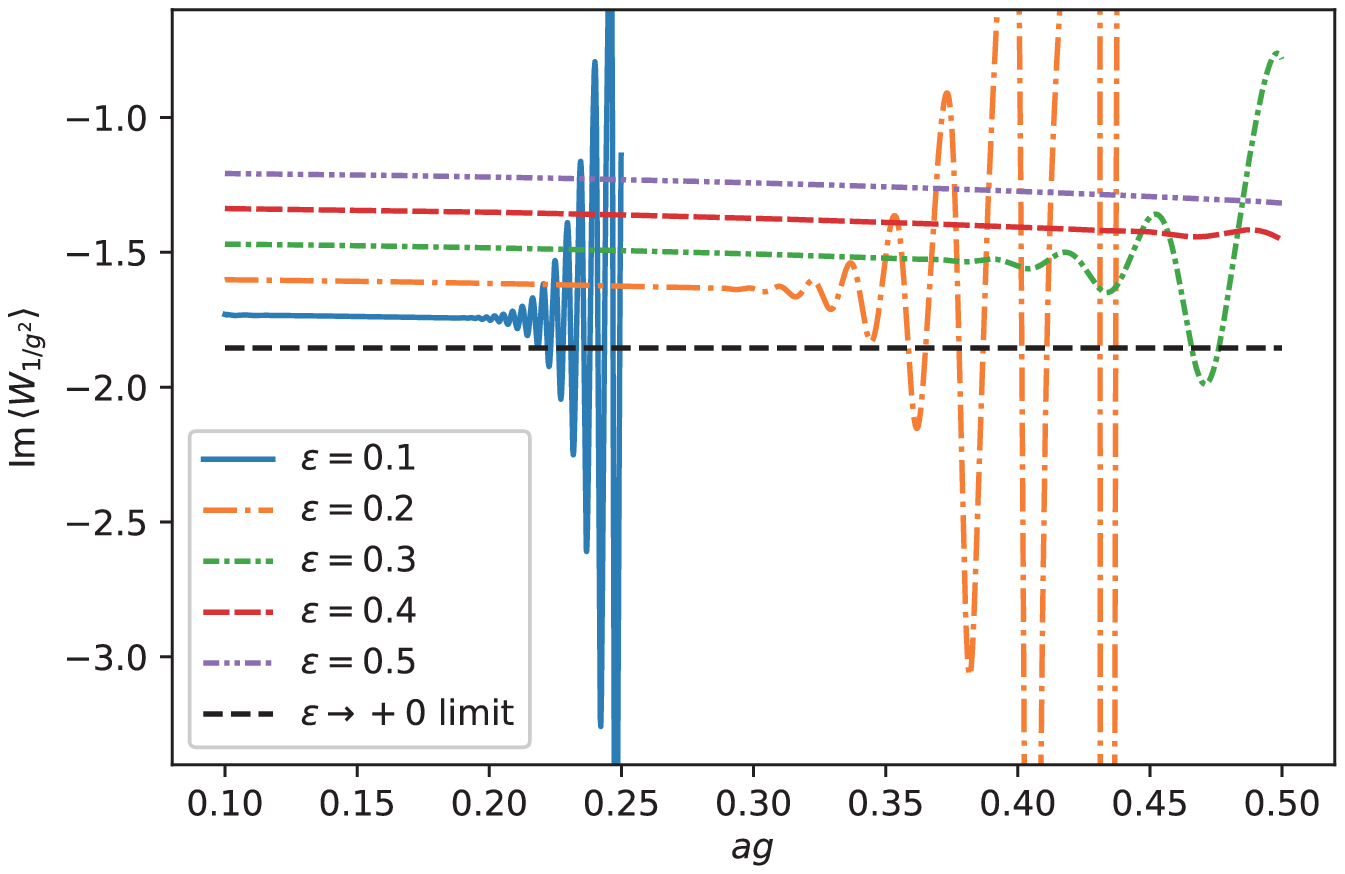} 
  \caption{
    \label{fig:re_w_su3}
    The expectation value of the Wilson loop $\langle W_{A} \rangle$
    with the area $A=1/g^2$
    evaluated with
    the analytic formula \eqref{eq:formula} for $SU(3)$.
    The values of $\varepsilon$ are varied to $\varepsilon = 0.1, \cdots, 0.5$.
    The plots are truncated before
    the curves become highly oscillatory.
    The black dashed line shows the
    $a\rightarrow 0$, $\varepsilon \rightarrow +0$ value,
    eq.~\eqref{eq:cont_string}.
  }
\end{figure}

\begin{figure}[htb]
  \centering
  \includegraphics[width=80mm]{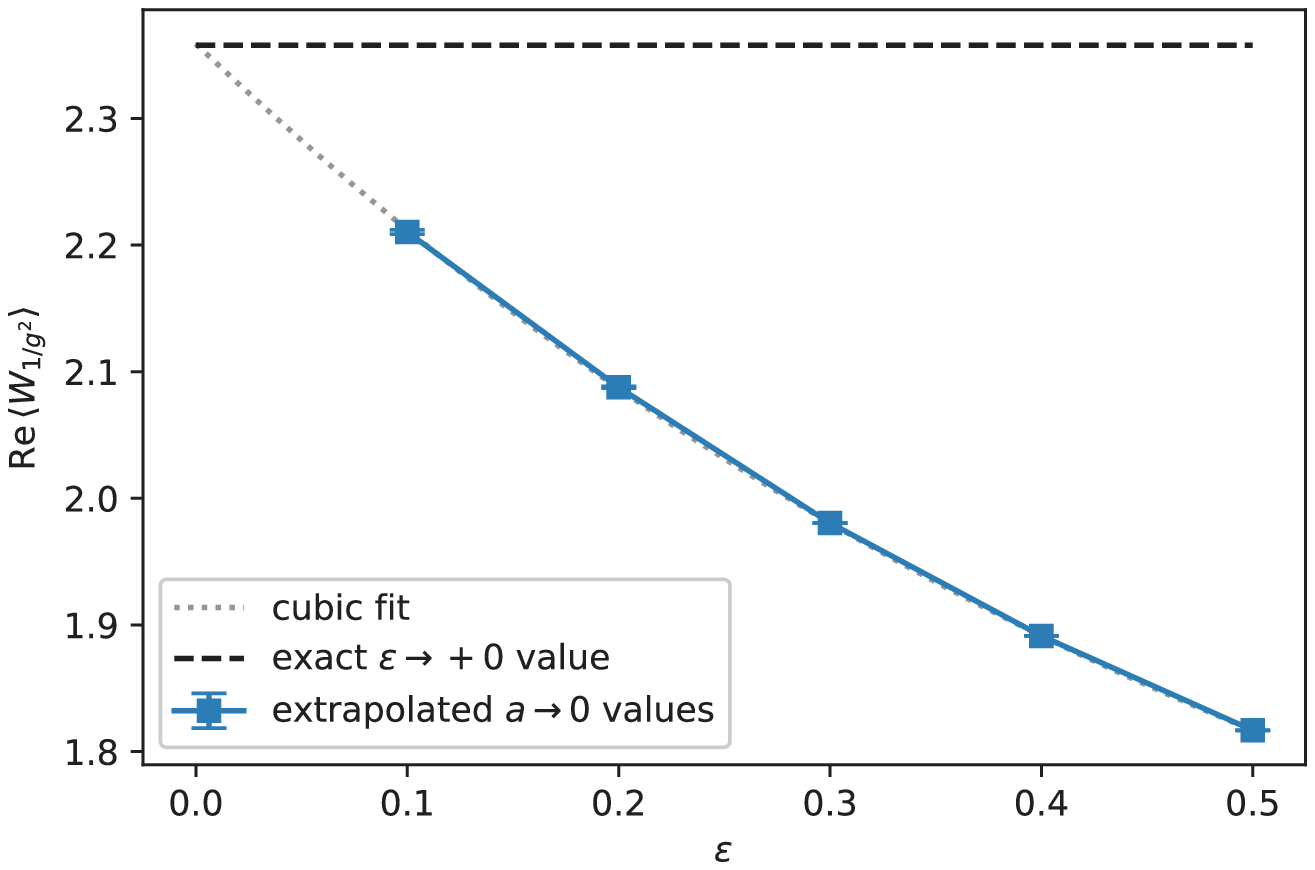}
  \vspace{5mm}
  \includegraphics[width=80mm]{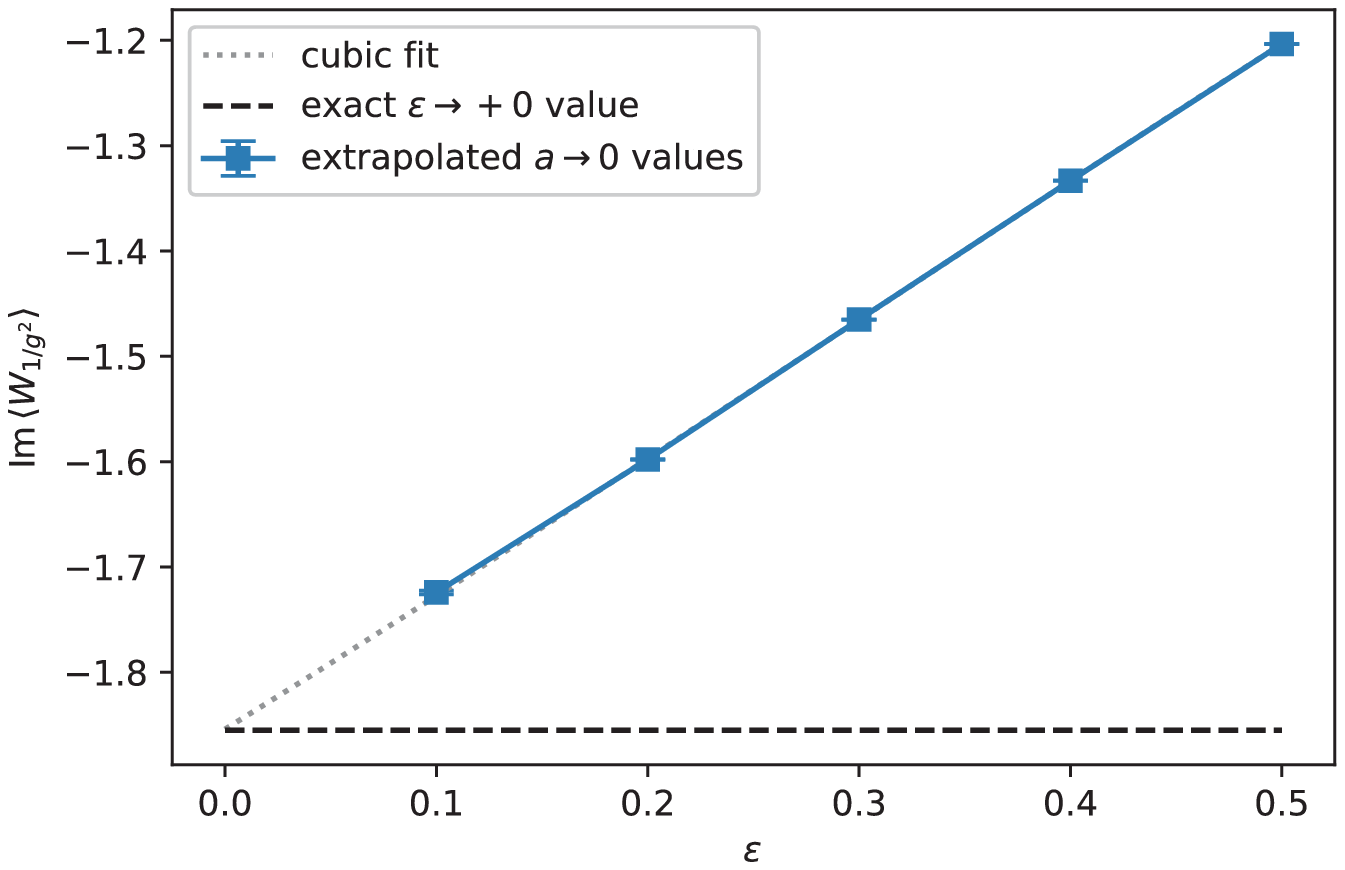} 
  \caption{
    The extrapolated $a\rightarrow0$ values of $\langle W_{1/g^2} \rangle$
    with various $\varepsilon$ for $SU(3)$.
    The $a\rightarrow0$ values are then fitted
    to obtain the final $\varepsilon\rightarrow +0$ result.
    The exact $\varepsilon\rightarrow +0$ value \eqref{eq:cont_string}
    is shown with the black dashed line for comparison.
    \label{fig:eps_extrapolation_su3}
  }
\end{figure}

For $SU(3)$,
we show in figure~\ref{fig:re_w_su3} the
expectation value $\langle W_{A} \rangle$ with the area $A=1$
and in figure~\ref{fig:eps_extrapolation_su3}
the extrapolation of the $a \rightarrow 0$ values
to the $\varepsilon \rightarrow +0$ limit.
\noindent
\noindent
The extrapolations are performed similarly to the $SU(2)$ case,
where we replace the range of $a$ to $a = 0.1, 0.125, \cdots, 0.2$.
The obtained estimate
$\lim_{a\rightarrow 0,\varepsilon\rightarrow +0} \langle W_{A=1} \rangle \approx 2.359(22) - 1.854(19)i$
agrees with the analytical value
$\lim_{a\rightarrow 0, \varepsilon\rightarrow +0} \langle W_{A=1} \rangle = 2.358 -1.855i$ within the error.
The chi-squared for the cubic fits are $\chi^2/{\rm DOF} = 3.3$ and $6.7$,
respectively, for the real and imaginary parts.
The above investigations show that the Wilson action with the $i\varepsilon$
correctly reproduces the appropriate continuum limit.%
\footnote{
  For the range of $\beta_t$ studied here,
  the asymptotic expansion
  do not converge up to machine precision
  in the calculation of the character coefficients in the $SU(3)$ case.
  The corresponding plot is therefore not shown in figure~\ref{fig:re_w_su3}.
}

\subsection{  $i\varepsilon$ in the derivation of the path integral}
\label{sec:operator}

In this subsection, we argue the reason why we need the $i\varepsilon$
for the Wilson action from the Hamiltonian formalism.
For this, we use the $SU(2)$ Wilson action as an example,
and follow the conventional Hamiltonian formalism
of the Wilson action \cite{Kogut:1974ag, Creutz:1976ch}.
To get rid of the complication related to the gauge symmetry,
we take the temporal gauge, $U_{x,t}=1$.
We keep the spatial lattice spacing $a$ finite in this subsection.
Then, at time slice $t$,
the degrees of freedom of the system are
the spatial link variables $U_{\textbf{x}, i}$.
To describe fluctuations around $U_{\textbf{x}, i}$,
we introduce the local coordinates $\theta_{\textbf{x}, i}^a$ by
\begin{align}
  e^{i \theta_{\textbf{x}, i}^a T^a} U_{\textbf{x}, i}.
\end{align}
In particular, we can track the infinitesimal time evolution
in terms of $\theta_{\textbf{x}, i}^a$.
With the conjugate momentum:
\begin{align}
  p_{\textbf{x}, i}^a \equiv \frac{a}{g^2}\dot{\theta}_{\textbf{x}, i}^a,
\end{align}
we can write down the Hamiltonian \cite{Kogut:1974ag}:
\begin{align}
  H \equiv \frac{g^2}{2a} \sum_{\textbf{x}, i} (p_{\textbf{x}, i}^a)^2 + V(U),
\end{align}
where we defined the potential:
\begin{align}
  V(U) \equiv
  \frac{2N_c}{a g^2}\sum_{\textbf{x}, i<j}
  \Big(1-\frac{1}{N_c}{\rm Re}\,{\rm tr}\,[U_{\textbf{x},i}U_{\textbf{x}+i,j}
  U_{\textbf{x}+j,i}^\dagger U_{\textbf{x},j}^\dagger]\Big).
\end{align}

We now derive the amplitude in path integral form for the $SU(2)$ Wilson theory.
The canonical operators 
$\hat{U}_{\textbf{x}, i}$, $\hat{p}_{\textbf{x}, i}^a$
satisfy the commutation relation:
\begin{align}
  [\hat{U}_{\textbf{x}, i}, \hat{p}_{\textbf{x}, i}^a] = T^a \hat{U}_{\textbf{x}, i}.
\end{align}
Configuration basis consists of the tensor product states:
\begin{align}
  \vert U \rangle \equiv \prod_{\textbf{x}, i} \vert U_{\textbf{x}, i} \rangle,
  \label{eq:U_field_state}
\end{align}
where
\begin{align}
  \hat{U}_{\textbf{x}, i} \vert U_{\textbf{x}, i} \rangle = U_{\textbf{x}, i} \vert U_{\textbf{x}, i} \rangle.
\end{align}
It is convenient to introduce another basis \cite{Chin:1985ua}:
\begin{align}
  \vert \{j_{\textbf{x}, i},m_{\textbf{x}, i},m_{\textbf{x}, i}'\} \rangle
  \equiv  \prod_{\textbf{x}, i} \vert j_{\textbf{x}, i},m_{\textbf{x}, i},m_{\textbf{x}, i}' \rangle,
\end{align}
where
\begin{align}
  \langle U_{\textbf{x}, i} \vert j,m,m' \rangle \equiv D_{m,m'}^j(U_{\textbf{x}, i})
\end{align}
with the matrix elements $D_{m,m'}^j(U)$ of the $SU(2)$ matrix $U$
in the spin $j$ representation.
From the Peter-Weyl theorem,
the basis $\vert \{j_{\textbf{x}, i},m_{\textbf{x}, i},m_{\textbf{x}, i}'\}\rangle$ satisfies the completeness relation:
\begin{align}
  1 = \sum_{\{j_{\textbf{x}, i},m_{\textbf{x}, i},m_{\textbf{x}, i}' \}}
  \Big( \prod_{\textbf{x}, i} (2j_{\textbf{x}, i}+1) \Big)
  \vert \{j_{\textbf{x}, i},m_{\textbf{x}, i},m_{\textbf{x}, i}' \} \rangle
  \langle \{j_{\textbf{x}, i},m_{\textbf{x}, i},m_{\textbf{x}, i}'\} \vert.
\end{align}
Furthermore, for finite $\eta_{\textbf{x}, i}^a$,
\begin{align}
  \langle U_{\textbf{x}, i}\vert
  e^{i\eta_{\textbf{x}, i}^a \hat{p}_{\textbf{x}, i}^a} | j, m,m '\rangle
  &= \big(T e^{i\int_0^1 ds\, \eta_{\textbf{x}, i}^a \mathcal{P}^a(s \eta_{\textbf{x}, i}) }
    \langle U_{\textbf{x}, i}\vert \big)
    \vert j,m,m'\rangle \nonumber\\
  &= \big[ T e^{i\int_0^1 ds\, \eta_{\textbf{x}, i}^a \mathcal{P}^a(s \eta_{\textbf{x}, i}) }
    D^j(U_{\textbf{x}, i})\big]_{m,m'},
\end{align}
where $T$ denotes the ordered product of the matrices
and $\mathcal{P}^a(\theta)$
are the differential operators expressed in terms of
the local coordinates on each link, $\theta^a$
\cite{Kogut:1974ag, Creutz:1976ch, Menotti:1981ry,
  Chin:1985ua}.
In particular, $-(\mathcal{P}^a(\theta))^2$
is the Laplacian on $S^3$
and $(\mathcal{P}^a(0))^2 = (i^{-1}\partial_{\theta^a})^2$. Thus,
\begin{align}
  \langle U_{\textbf{x}, i}\vert
  (\hat{p}_{\textbf{x}, i}^a)^2 | j, m,m '\rangle
  =
  [(\mathcal{P}^a(0))^2 D^j(U_{\textbf{x}, i})]_{m,m'}
  =j(j+1)D^j_{m,m'}(U_{\textbf{x}, i}).
\end{align}

We now calculate the amplitude from the state $\psi_i$ to $\psi_f$:
\begin{align}
  A_{\psi_f,\psi_i}(T) \equiv \langle \psi_f \vert e^{-i\hat{H} T}\vert \psi_i \rangle.
\end{align}
We discretize $T\equiv N a_0$ and ignore higher order terms of $a_0$.
Note that
\begin{align}
  \langle U' \vert e^{-i a_0 \hat{H}} \vert U \rangle
  &=
    \langle U' \vert
    e^{-i \frac{a_0 g^2}{2a}\sum_{\textbf{x}, i} (\hat{p}_{{\textbf{x}, i}}^a)^2}
    e^{-i a_0 V(\hat{U})} \vert U \rangle \nonumber\\
  &=
    \prod_{\textbf{x}, i}
    \Big[
    \sum_{j_{\textbf{x}, i}}
    (2j_{\textbf{x}, i}+1)
    \chi_{j_{\textbf{x}, i}} (U'_{{\textbf{x}, i}} U_{{\textbf{x}, i}}^\dagger)
    e^{-i\frac{a_0 g^2}{2a} j_{\textbf{x},i}(j_{\textbf{x},i}+1)}
    \Big]
    e^{-i a_0 V(U)} .
    \label{eq:infinitesimal}
\end{align}
By diagonalizing
\begin{align}
  U'_{{\textbf{x}, i}} U_{{\textbf{x}, i}}^\dagger
  \sim
  {\rm diag} (e^{i\delta\phi_{\textbf{x}, i}}, e^{-i\delta\phi_{\textbf{x}, i}})
  \quad \big( \delta \phi_{\textbf{x}, i} \in [-\pi, \pi) \big),
\end{align}
we can write the expression in the bracket
appearing in eq.~\eqref{eq:infinitesimal} as
(we drop the subscripts ${\textbf{x}, i}$ temporarily
for notational simplicity)
\begin{align}
  &\sum_{j} (2j+1)
    \frac{\sin (2j+1)\delta\phi}{\sin \delta\phi}
    e^{-i \frac{a_0 g^2}{2a} j(j+1)} \nonumber\\
  &= -\frac{1}{2}
    \frac{1}{\sin \delta\phi}
    e^{i \frac{a_0 g^2}{8a}}
    \frac{d}{d\delta\phi}
    \sum_{n \geq 1 }
    \Big[
    e^{-i\frac{a_0 g^2}{8a} n^2 + in \delta\phi}
    +
    e^{-i\frac{a_0 g^2}{8a} n^2 - in \delta\phi}
    \Big] \nonumber\\
  &= -\frac{1}{2}
    \frac{1}{\sin \delta\phi}
    e^{i \frac{a_0 g^2}{8a}}
    \frac{d}{d\delta\phi}
    \vartheta\Big(\frac{\delta \phi}{2\pi}, -\frac{a_0 g^2}{8\pi a}\Big),
    \label{eq:square_bracket}
\end{align}
where we defined $n \equiv 2j+1 $ in the second line.
In order to further rewrite the expression,
we introduce an infinitesimal imaginary part:
\begin{align}
  \vartheta\Big(\frac{\delta \phi}{2\pi}, -\frac{a_0 g^2}{8\pi a}\Big)
  \to
  \vartheta\Big(\frac{\delta \phi}{2\pi}, -e^{-i \varepsilon}\frac{a_0 g^2}{8\pi a}\Big)
\end{align}
The resulting function has a sharp peak around $\delta\phi = 0$, and thus
\begin{align}
  &-\frac{1}{2}
  \frac{1}{\sin \delta\phi}
    e^{i \frac{a_0 g^2}{8a}}
    \frac{d}{d\delta\phi}
    \vartheta\Big(\frac{\delta \phi}{2\pi}, -e^{-i \varepsilon}\frac{a_0 g^2}{8\pi a}\Big)
    \nonumber\\
  &\approx
  -\frac{1}{2}
  \frac{1}{\sin \delta\phi}
    e^{i \frac{a_0 g^2}{8a}}
    \frac{d}{d\delta\phi}
    e^{i\pi/4} \sqrt{\frac{8\pi a}{-e^{-i \varepsilon} a_0 g^2}}
    \exp\Big[ie^{i \varepsilon} \frac{2a}{a_0 g^2}(\delta \phi)^2\Big]\nonumber\\
  &=
    {\rm const}\cdot
    \frac{\delta\phi}{\sin \delta\phi}
    \exp\Big[ie^{i \varepsilon} \frac{2a}{a_0 g^2}(\delta \phi)^2\Big],
  \label{eq:mid_eq}
\end{align}
where in the second line
we dropped the contributions
with nontrivial winding that will be exponentially suppressed
in the $a_0 \rightarrow 0$ limit.
Finite contribution comes from the fluctuations of order
$\delta\phi = O(a_0)$.
With the similar argument as in eq.~\eqref{eq:nonlinearlity},
we can rewrite eq.~\eqref{eq:mid_eq} up to an overall constant as
\begin{align}
  {\rm const}\cdot \frac{\delta\phi}{\sin \delta\phi}
  \exp\Big[ i e^{i \varepsilon} \frac{2a}{a_0 g^2} (\delta\phi)^2\Big]
  ={\rm const'} \cdot
  \exp\Big[ -i e^{i \varepsilon} \frac{2a}{a_0 g^2} {\rm tr}\, [U' U^\dagger] \Big].
\end{align}
The amplitude is thus
rewritten with the plaquette action
in the desired path integral form
with the $i\varepsilon$:
\small
\begin{align}
  &A_{n_f,n_i}(T) \approx
    {\mathcal N}'
    \lim_{\varepsilon \rightarrow + 0}
    \int \Big(\prod_{\ell=0}^N dU_\ell\Big)
    \exp\Big[ i \sum_{\ell=0}^{N-1}
    \Big\{
    -e^{i\varepsilon}\frac{2a}{a_0 g^2}
    \sum_{\textbf{x}, i}
    {\rm tr}
    [U_{\ell+1,\textbf{x}, i} U_{\ell\textbf{x}, i}^\dagger] \nonumber\\
  &~~~~~~~~~~~~~~~~~~~~~~~~~~~~~~~~~
    + \frac{2a_0}{a g^2}
    \sum_{\textbf{x}, i<j}{\rm tr}
    \,[U_{\ell,\textbf{x},i}U_{\ell,\textbf{x}+i,j}
    U_{\ell,\textbf{x}+j,i}^\dagger U_{\ell,\textbf{x},j}^\dagger]
    \Big\}
    \Big] \psi_f^*(U_N) \psi_i(U_0),
\end{align}
\normalsize
where ${\mathcal N}'$ is a normalization constant.

Despite the complications related to the field theory,
the basic structure is the same as the
quantum mechanical model in section~\ref{sec:qm}.
The Wilson action is only guaranteed to reproduce the continuum action for smooth fields,
and we need to suppress the contributions from
singular paths in advance with the $i\varepsilon$.
The $i\varepsilon$ should thus be regarded as
a part of the definition of the real-time path integral
when using the Wilson action.

Note also that, since we only have considered the formal $a_0\rightarrow 0$ limit,
the $i\varepsilon$ in the spatial plaquettes have not appeared in the discussion.
In fact, in this treatment, the characters for the spatial plaquettes
can be expressed in terms of the modified Bessel function of the form
$ I_n ( 2i a_0/(a g^2))$
[see eq.~\eqref{eq:charac_coeff}],
for which we can apply the expansion of $I_n(z)$ around zero:
\begin{align}
  I_n (z) = \Big(\frac{z}{2}\Big)^n \sum_{k\geq 0}  \frac{(z/2)^{2k}}{n! (n+k)!}.
  \label{eq:expansion_zero}
\end{align}
The characters
coming from the spatial plaquettes are
thus analytic in the limit $a_0 \rightarrow 0$
for a fixed $a$,
giving no complication.
The subtlety for the spatial plaquettes arises
when we take the
continuum limit taking $a_0 \rightarrow 0$ and $a \rightarrow 0$
at the same time,
making $g^2$ run according to the renormalization group equation.
In the latter treatment,
which is required in extracting the continuum physics,
we need to incorporate $i\varepsilon$ also for the spatial plaquettes
as argued in subsection~\ref{sec:i_eps_lattice}.

\section{
  Summary and outlook}
\label{sec:summary}

In this paper, we discussed that the $i\varepsilon$ is an
essential ingredient in defining the real-time path integral
for the Wilson action,
and showed how its necessity can be explained
from the Hamiltonian formalism.
In numerical calculations,
one needs to take the $i\varepsilon$ into account
both for the timelike and spacelike plaquettes, 
and this can be done by calculating the continuum limit with several $\varepsilon$
and taking the $\varepsilon \rightarrow +0$ limit,
or rewriting the Boltzmann weight in terms of characters
dropping the unwanted part
of the asymptotic expansion of modified Bessel function
for the character coefficients.
We in particular demonstrated that,
with the $i\varepsilon$,
the Wilson action gives the correct continuum limit
using the two-dimensional theory as an example.
We believe that this clarification of the subtlety
helps us investigate more involved cases
such as full quantum chromodynamics.

As we commented in section~\ref{sec:convergence},
we need to choose $\varepsilon$ large enough
for a given lattice spacing to avoid the oscillation
coming from the unwanted part of the asymptotic expansion.
For the studied range of the lattice spacing,
this is satisfied numerically in two-dimension at
$\beta_t \sin \varepsilon \gtrsim 4.5$ for $SU(2)$
and
$\beta_t \sin \varepsilon \gtrsim 15$ for $SU(3)$.
Since the characters are expressed with
$\beta_r$ in eq.~\eqref{eq:charac_coeff},
these values should give a rough estimate
of the required $\varepsilon$
also in higher dimensions.
The rather large bounds are, however,
unpleasant for the four-dimensional application
because of the existence of the
critical slowing down at large $\beta_r$.
A similar situation occurs for
the action expressed with the characters
in which the modified Bessel function is replaced
by its asymptotic expansion dropping the unwanted part.
This is because the asymptotic expansion itself is divergent,
and thus we need to 
choose the order to truncate the expansion.
For large enough $\beta_r$,
the summand becomes smaller than
the machine precision at some order,
and thus we can truncate the expansion there.
However, comparably large $\beta_r$ is required
for such convergence especially in the $SU(3)$ case.
Therefore,
though our method gives
a way to obtain the appropriate continuum prediction,
it is desirable to circumvent the
critical slowing down (see, e.g.,
\cite{Parisi:1984,Batrouni:1985jn,Davies:1987vs,Katz:1987ti,
  Davies:1989vh,Luscher:2009eq,Luscher:2011kk,Albergo:2019eim,
  Foreman:2021ixr,Albandea:2021lvl,Foreman:2021ljl,Nguyen:2021zgx})
or develop an action that is convergent at small $\beta_r$
by, e.g., the contour deformation \cite{Kanwar:2021tkd}.
Studies along these lines are in progress and will be reported elsewhere.

\section*{Acknowledgments}
The author thanks
Andrei Alexandru, 
Michael Austin DeMarco,
Shoji Hashimoto,
Taku Izubuchi, Luchang Jin, Scott Lawrence,
Jun Nishimura,
Iso Satoshi, 
and Akio Tomiya
for valuable discussions.
In particular, he thanks Taku Izubuchi
for sharing a Monte Carlo code
and Akio Tomiya for introducing the software LatticeQCD.jl,
which helped a part of the study in section~\ref{sec:convergence}.
The author is further grateful to Norman Christ for
the discussions in the early stage of the study,
to Masafumi Fukuma and Yusuke Namekawa
for giving important comments on the manuscript,
and to Yoshio Kikukawa and the referee of
Progress of Theoretical and Experimental Physics
for pointing out the misstatements in
the argument of section~\ref{sec:explanation}
in the first version of the manuscript.
This work is supported by the Special Postdoctoral Researchers Program 
of RIKEN and by
JSPS KAKENHI Grant Number JP22H01222.


\baselineskip=0.9\normalbaselineskip



\end{document}